\documentclass[useAMS,usenatbib]{mn2e}

\usepackage{epsfig}
\usepackage{amsmath}
\usepackage{amssymb}
\usepackage{natbib}
\usepackage{threeparttable} 
\usepackage{booktabs}

\usepackage{epstopdf}
\usepackage{times}

\newcommand{\chan}{\textit{Chandra}}
\newcommand{\swift}{\textit{Swift}}
\newcommand{\rxte}{\textit{RXTE}}

\newcommand{\inte}{\textit{Integral}}
\newcommand{\maxi}{\textit{MAXI}}

\newcommand{\rosat}{\textit{ROSAT}}

\newcommand{\nustar}{\textit{NuSTAR}}

\newcommand{\Msun}{\mathrm{M}_{\odot}}
\newcommand{\lum}{\mathrm{erg~s}^{-1}}
\newcommand{\flux}{\mathrm{erg~cm}^{-2}~\mathrm{s}^{-1}}

\newcommand{\cnts}{\mathrm{c~s}^{-1}}

\newcommand{\nh}{\mathrm{cm}^{-2}}
\newcommand{\dist}{(D/5.8~\mathrm{kpc})^2}

\newcommand{\kms}{\mathrm{km~s}^{-1}}
\newcommand{\gmc}{GM/c^2}
\newcommand{\risco}{R_{\mathrm{ISCO}}}

\newcommand{\rg}{R_{\mathrm{g}}}
\newcommand{\dens}{\mathrm{cm}^{-3}}

\newcommand{\source}{RXS~J1804}
\newcommand{\sourcelong}{1RXS~J180408.9--342058}

\def \nar {NewAR}

\def \mnras {MNRAS}
\def \apj {ApJ}
\def \apjs {ApJS}
\def \apjl {ApJL}
\def \aap {A\&A}
\def \nat {Nature}
\def \araa {ARAA}
\def \atel {ATel}

\def \pasj {PASJ}

\def \procspie {Proc. SPIE}

\def \apss {Ap\&SS}

\title[Accretion disk in \sourcelong]{Disk reflection and a possible disk wind during a soft X-ray state in the neutron star low-mass X-ray binary \sourcelong}
\author[N. Degenaar et al.]
{N. Degenaar$^{1,2}$\thanks{e-mail: degenaar@ast.cam.ac.uk}, D.~Altamirano$^3$, M. Parker$^{1}$, J.C.A.~Miller-Jones$^{4}$, J.M.~Miller$^{5}$, 
\newauthor C.O.~Heinke$^6$, R. Wijnands$^{2}$, R. Ludlam$^{5}$, A. Parikh$^{2}$, J.W.T. Hessels$^{2,7}$, N. Gusinskaia$^{2}$, 
\newauthor A.T. Deller$^7$  and A.C. Fabian$^{1}$ \\
$^1$Institute of Astronomy, University of Cambridge, Madingley Road, Cambridge CB3 OHA, UK\\
$^2$Anton Pannekoek Institute for Astronomy, University of Amsterdam, Science Park 904, 1098 XH, Amsterdam, the Netherlands\\
$^3$School of Physics and Astronomy, University of Southampton, Southampton, Hampshire, SO171BJ, UK\\
$^4$International Centre for Radio Astronomy Research, Curtin University, GPO Box U1987, Perth, WA 6845, Australia\\
$^5$Department of Astronomy, University of Michigan, 1085 South University Avenue, Ann Arbor, MI  48109, USA\\
$^6$Department of Physics, University of Alberta, 4-183 CCIS, Edmonton, AB T6G 2E1, Canada\\
$^7$ASTRON, the Netherlands Institute for Radio Astronomy, Postbus 2, 7990 AA, Dwingeloo, The Netherlands
}

\begin{document}

\date{Accepted to MNRAS Main Journal}

\pagerange{\pageref{firstpage}--\pageref{lastpage}} \pubyear{0000}

\maketitle

\label{firstpage}

\begin{abstract}
\sourcelong\ is a transient neutron star low-mass X-ray binary that exhibited a bright accretion outburst in 2015. We present \nustar, \swift, and \chan\ observations obtained around the peak brightness of this outburst. The source was in a soft X-ray spectral state and displayed an X-ray luminosity of $L_{\mathrm{X}}$$\simeq$$(2-3)\times10^{37}~\dist~\lum$ (0.5--10 keV). 
The \nustar\ data reveal a broad Fe-K emission line that we model as relativistically broadened reflection to constrain the accretion geometry. We found that the accretion disk is viewed at an inclination of $i$$\simeq$$27^{\circ}$--$35^{\circ}$ and extended close to the neutron star, down to $R_{\mathrm{in}}$$\simeq$5--7.5 gravitational radii ($\simeq$11--17~km). This inner disk radius suggests that the neutron star magnetic field strength is $B$$\lesssim$$2\times10^8$~G. We find a narrow absorption line in the \chan/HEG data at an energy of $\simeq$7.64~keV with a significance of $\simeq$4.8$\sigma$. This feature could correspond to blue-shifted Fe\,{\sc xxvi} and arise from an accretion disk wind, which would imply an outflow velocity of $v_{\mathrm{out}}$$\simeq$$0.086c$ ($\simeq$25\,800$~\kms$). However, this would be extreme for an X-ray binary and it is unclear if a disk wind should be visible at the low inclination angle that we infer from our reflection analysis. Finally, we discuss how the X-ray and optical properties of \sourcelong\ are consistent with a relatively small ($P_{\mathrm{orb}}$$\lesssim$3~hr) binary orbit.
\end{abstract}


%%%%%%%%%%%%%%%%%
% INTRODUCTION
%%%%%%%%%%%%%%%%%

\section{Introduction}\label{sec:introduction}
Low-mass X-ray binaries (LMXBs) consist of a neutron star or a black hole plus a companion star that is less massive than the compact primary. The companion typically overflows its Roche-lobe, thereby feeding gas to an accretion disk that surrounds the neutron star or black hole. In transient systems, outbursts of active accretion are interleaved with long periods of quiescence when the accretion rate onto the compact primary is strongly reduced. LMXBs are excellent laboratories to investigate how accretion proceeds under the influence of strong gravity over a wide range of accretion rates. 

X-ray spectroscopy is a powerful tool to study accretion processes in LMXBs; both the X-ray continuum spectral shape and discrete emission/absorption features give insight into the accretion geometry. Broadly speaking, two main X-ray spectral states can be distinguished during the active phases of LMXBs \citep[see e.g.,][for a detailed overview]{homan2005_specstates,remillard2006}. During a ``soft state'', thermal emission from the accretion disk is prominently detected in the X-ray spectrum. However, during a ``hard state'', the disk emission is much weaker and a strong non-thermal, hard X-ray emission component is seen that likely arises from a hot plasma in the inner accretion flow.

Hard X-rays can illuminate the accretion disk and produce a reflection spectrum of emission lines, the most prominent being Fe-K at 6.4--6.97~keV \citep[e.g.,][]{george1991,matt1991}. The shape of such lines depends on the chemical abundances and ionization state of the accretion disk, but also on its inner radial extent since both relativistic and dynamical effects act to broaden reflected emission lines. Disk reflection studies suggest that in some neutron star LMXBs the stellar magnetic field is strong enough to truncate the inner accretion disk \citep[][]{degenaar2014_groj1744,pintore2016,king2016}, whereas in others it can extend all the way down to the innermost stable circular orbit (ISCO) or the neutron star surface \citep[e.g.,][]{bhattacharyya2007,dai2009,papitto2009,cackett2010_iron,miller2013_serx1,degenaar2015_4u1608,disalvo2015}.

High-resolution X-ray spectroscopy of LMXBs can reveal narrow, blue-shifted absorption features  arising from a highly ionized wind that is blown off the accretion disk \citep[e.g.,][]{brandt2000,ueda2004,miller2006,miller2016_wind,diaztrigo2007,bozzo2016_wind}. These winds are preferentially detected during soft X-ray spectral states \citep[e.g.,][]{miller2006_winds,miller2008,neilsen2009,ponti2012_winds}. Typical outflow velocities are $\simeq$400--3000$~\kms$ ($\simeq$0.001--0.01$c$) and the properties are very similar in black hole and neutron star LMXBs \citep[e.g.,][]{diaztrigo2015}. Powerful disk winds may carry away a significant amount of mass, similar to or even exceeding the amount of mass that is accreted \citep[e.g.,][]{lee2002,neilsen2011,king2012,king2015,ponti2012_winds,degenaar2014_groj1744}. Winds are therefore important in considering the energy budget of the accretion process, and may also lead to instabilities in the accretion flow \citep[e.g.,][]{begelman1983b,shields1986}.

\vspace{-0.2cm}
\subsection{\sourcelong\ (\source)}\label{subsec:source}
\sourcelong, hereafter referred to as \source, was detected as a faint X-ray source with \rosat\ in 1990 \citep[][]{voges1999}. It remained unclassified until 2012, when \inte\ detected a thermonuclear X-ray burst that identified \source\ as an accreting neutron star in, most likely, an LMXB. Assuming that the peak luminosity of the X-ray burst did not exceed the empirically-determined Eddington limit \citep[$L_{\mathrm{Edd}}$$=$$3.8\times10^{38}~\lum$;][]{kuulkers2003}, places the source at a distance of $D$$\lesssim$5.8~kpc \citep[][]{chenevez2012}. 

No persistent emission could be detected with \inte\ when the X-ray burst was seen in 2012, implying an accretion luminosity of $L_{\mathrm{X}}$$\lesssim 4 \times 10^{35}~\lum$ at the time (3--10~keV). Indeed, follow-up \swift\ observations detected the source at $L_{\mathrm{X}}$$\simeq$$10^{33}-10^{34}~\lum$ (0.5--10~keV), suggesting that it had exhibited a faint accretion outburst that was missed by all-sky monitors \citep[][]{chenevez2012,kaur2012_1804}. 

\begin{figure}
 \begin{center}
	\includegraphics[width=8.5cm]{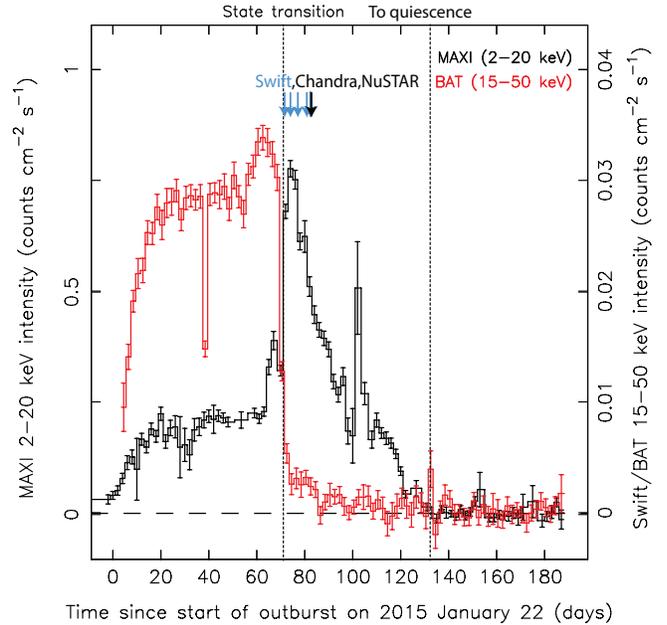}
    \end{center}
    \caption[]{Combined \swift/BAT (red, 15--50 keV) and \maxi\ (black, 2--20 keV) X-ray light curves of the 2015 outburst of \source\ (2-day bins). The hard to soft state transition and the estimated return to quiescence are indicated by the vertical dotted lines. Times of the \chan, \nustar\ and \swift/XRT observations discussed in this work are indicated by the arrows. 
        }
 \label{fig:bat}
\end{figure}

In 2015 January, an outburst from \source\ was seen by \swift/BAT and \maxi\ monitoring observations \citep[][]{krimm2015,negoro2015}. Figure~\ref{fig:bat} shows the publicly available \maxi\ \citep[2--20~keV;][]{maxi2009}\footnote{http://maxi.riken.jp/top/index.php?cid=1\&jname=J1804-343} and \swift/BAT \citep[15--50 keV;][]{krimm2013}\footnote{http://swift.gsfc.nasa.gov/results/transients/weak/1RXSJ180408.9-342058/} daily monitoring light curves of this outburst. The source remained at relatively constant flux in a hard X-ray spectral state \citep[][]{ludlam2016} for $\simeq$2~months, during which radio emission from a jet was detected \citep[][]{deller2015_1804}. Around April 3, however, the source transitioned to a soft X-ray spectral state \citep[][dotted line in Figure~\ref{fig:bat}]{degenaar2015_1804}. The activity ceased $\simeq$2~months later, i.e., the total outburst duration was $\simeq$4.5~months (see Figure~\ref{fig:bat}). 

In this work we report on X-ray spectral observations of \source\ obtained with \nustar\ \citep[][]{harrison2013_nustar}, \swift\ \citep[][]{gehrels2004}, and \chan\ \citep[][]{weisskopf2000}. These data were taken during the soft X-ray spectral state near the peak brightness of the 2015 outburst (see Table~\ref{tab:obs} and Figure~\ref{fig:bat}). 

%%%%%%%%%%%%%%%%%
% OBSERVATIONS
%%%%%%%%%%%%%%%%%

\section{Observations and data analysis}\label{sec:obs}

\subsection{\nustar}\label{subsec:nustar}
\source\ was observed with \nustar\ between 12:11 UT on 2015 April 14 and 00:51 UT on April 15 (Obs ID 90101003002), i.e. $\simeq$9~d after the transition to the soft spectral state (see Figure~\ref{fig:bat}). \nustar\ consists of two co-aligned grazing incidence hard X-ray imaging telescopes, the focal plane mirror (FPM) A and B, which provide sensitivity in an energy range of 3--79~keV. As such, it is a particularly powerful instrument to study disk reflection spectra.

Standard screening and processing with \textsc{nustardas}, incorporated in \textsc{heasoft} (v. 6.17), resulted in $\simeq$20~ks of on-target exposure time per mirror. We employed \textsc{nuproducts} to create light curves and spectra for the FPMA and FPMB. A circular region with a radius of $60''$ was used to extract source events and a void region of the same size, placed on the same CCD, was used to extract background events. The summed FPMA/B count rate varied between $\simeq$180--230$~\cnts$ during the observation (3--79 keV).

All spectral data were grouped into bins with a minimum of 20 photons using \textsc{grppha}. We found that \source\ was detected significantly above the background up to $\simeq$35~keV, so we restricted our spectral fits of the \nustar\ data to an energy range of 3--35~keV. The separate spectra obtained for the two mirrors were fitted simultaneously with all parameters tied, but with a constant factor varying between them to account for calibration uncertainties.

\begin{table}
\caption{Log of the (soft state) observations considered in this work.}
\begin{threeparttable}
\begin{tabular*}{0.49\textwidth}{@{\extracolsep{\fill}} l c c c}
\hline 
Instrument & ObsID & Start time  & Exposure time \\ %& Start time
&   & (MJD)  & (ks) \\ % & (UT)
\hline
\swift\ (XRT/WT) & 32436029 & 57115.83  & 1.1 \\ %& 2015 Apr 03 20:01
\swift\ (XRT/WT)  & 32436030 & 57118.18  & 1.0 \\ %& 2015 Apr 06 04:16
\swift\ (XRT/WT)  & 32436031 & 57121.17  & 1.0 \\ %& 2015 Apr 09 04:06
\swift\ (XRT/WT)  & 32436032 & 57124.16  & 1.0 \\ %& 2015 Apr 12 03:57
\nustar\ & 90101003002 & 57126.51  & 20.2 \\ %& 2015 Apr 14 12:11
\swift\ (XRT/WT)  & 81451001 & 57126.54  & 1.0 \\ %& 2015 Apr 14 13:03
\chan\ (HETG) & 17649 & 57126.62 & 30.0 \\ % & 2015 Apr 14 14:55
\hline
\end{tabular*}
\label{tab:obs}
\end{threeparttable}
\end{table}

\subsection{\chan}\label{subsec:chandra}
We observed \source\ with \chan\ on 2015 April 14 between 14:55 and 23:54 UT (Obs ID 17649) for $\simeq$30~ks. The High Energy Transmission Grating \citep[HETG;][]{canizares2000} was used to disperse the incoming light onto the ACIS-S array, enabling high X-ray spectroscopy to search for narrow spectral features. We focus on the data from the High Energy Grating (HEG) arm, since it has better energy resolution and extends to higher energies than the Medium Energy Grating (MEG), and is therefore best suited for studying highly ionized disk winds. The ACIS CCD was operated in continuous clocking (CC) mode to mitigate the effects of pile-up. Furthermore, a 100-column ``gray window'' was placed over the aimpoint to ensure that only one in every ten zeroth-order event was read out, to limit telemetry saturation and loss of frames.

We reduced the \chan\ data using \textsc{ciao} (v. 4.8). We followed the standard steps for CC-mode data reduction\footnote{http://cxc.harvard.edu/ciao/why/ccmode.html}, except that we ran \textsc{tg$\_$create$\_$mask} with a width factor of 15, rather than the default 35. We used this narrower extraction region in an attempt to reduce noise. \source\ was detected at a HEG count rate of $\simeq$50--60$~\cnts$ (0.8--10 keV). We did not find significant differences between the spectra of the plus and minus grating orders such as e.g., discussed in \citet{chiang2015}. We therefore combined the first-order positive and negative grating data using \textsc{combine$\_$grating$\_$spectra}. To avoid instrumental artifacts present in CC mode data of bright sources, we restricted the \chan/HEG spectral fits to 2--9~keV \citep[see e.g.,][]{cackett2009_ironchan,miller2011,miller2012_winds,chiang2015}. 

\subsection{\swift}\label{subsec:swift}
To follow the global X-ray spectral and flux evolution, the outburst of \source\ was monitored using many short ($\simeq$1--2~ks) pointed \swift\ observations, providing 0.3--10 keV coverage with the XRT \citep[e.g.,][]{degenaar2015_1804,deller2015_1804,krimm2015_2,baglio2016}. Here, we aim to characterize the tentative high-energy absorption features reported by \citet{degenaar2015_1804}. We therefore focus on five soft-state observations obtained around the peak of the outburst between 2015 April 3--14 (Table~\ref{tab:obs}). Observation 81451001 was performed simultaneously with our \nustar\ observation, and we used this data for broad-band spectral fitting. 

The \swift\ data were analyzed using standard tools incorporated in \textsc{heasoft} (v. 6.17). All observations considered in this work were obtained in the windowed timing (WT) mode. We used \textsc{XSelect} to extract source events from a box-shaped region with a length of $70''$. Background events were obtained from the outer regions of the CCD, avoiding the inner $280''$ around the source. In all observations considered here, the XRT count rate was $>$100$~\cnts$ and hence sufficiently high to cause pile-up. Following the recommended procedure, we therefore excised as many inner pixels as needed for the spectrum to remain unchanged after removing more pixels \citep[e.g.,][]{romano2006}. This resulted in the exclusion of the inner $2''$ for all five XRT observations.

Exposure maps were used to create ancillary response files with \textsc{xrtmkarf}, to account for hot pixels and bad columns on the CCD. The latest response matrix file was sourced from the calibration data base (v. 15). Spectra were grouped to contain a minimum of 20 photons per bin and spectral fits were performed in an energy range of 0.7--10 keV to avoid instrumental artifacts present for bright sources observed in WT mode.\footnote{See http://www.swift.ac.uk/analysis/xrt/digest cal.php.}

\subsection{Spectral analysis and assumed quantities}\label{subsec:specana}
All spectral fits reported in this work were performed within \textsc{XSpec} \citep[v. 12.9;][]{xspec}.  
For each fit we included the \textsc{tbabs} model to account for interstellar absorption along the line of sight. Abundances were set to \textsc{wilm} \citep[][]{wilms2000}, and cross-sections to \textsc{vern} \citep[][]{verner1996}. The total model 0.7--35 keV flux and error were determined with the \textsc{cflux} convolution model. When fitting spectral data from different instruments together, we used a constant multiplication factor ($C$) to allow for cross-calibration uncertainties. Throughout this work, we assume a source distance of $D$$=$5.8~kpc, an Eddington limit of $L_{\mathrm{Edd}}$$=$$3.8\times10^{38}~\lum$ \citep[][]{kuulkers2003}, and a neutron star mass of $M$$=$1.5~$\Msun$. Errors are given at 1$\sigma$ confidence level.

 \begin{figure}
 \begin{center}
	\includegraphics[width=8.5cm]{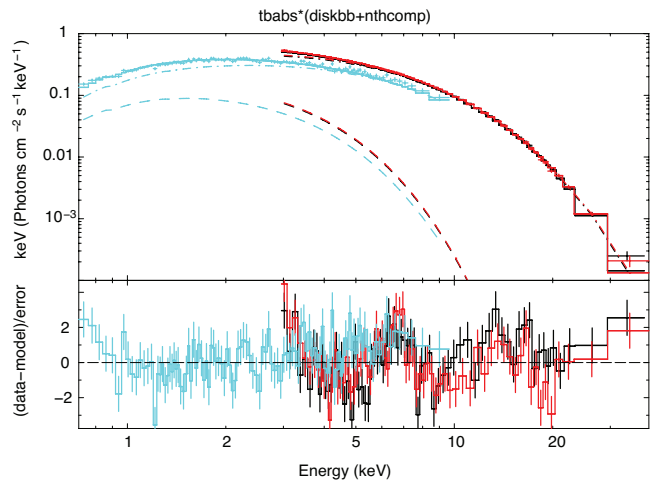}
    \end{center}
    \caption[]{\nustar\ (FPMA in black, FPMB in red) and \swift/XRT (cyan) unfolded spectra (rebinned for visual clarity). Solid curves represent a fit to a continuum model consisting of a disk black body (dashed curves), and Comptonized emission (dashed-dotted curves). The bottom panel shows the residuals in sigmas, revealing un-modeled structure near 6--7~keV that likely also causes deviations from the model at lower and higher energies.
        }
 \label{fig:contspec}
\end{figure}

%%%%%%%%%%%%%%%%%
% SPECTRA
%%%%%%%%%%%%%%%%%

\section{Results}\label{sec:results}

\subsection{Broad-band X-ray spectrum in the soft state}\label{subsec:broad}
The \nustar\ observation of \source\ was carried out simultaneously with the \chan/HEG observation and the \swift/XRT observation with ID 81451001 (see Table~\ref{tab:obs}). Since the \swift\ data are more reliable to lower energies than our \chan\ data (see Section~\ref{sec:obs}), we choose to use the \swift\ observation together with the \nustar\ data to perform broad-band spectral fits. 

We fitted the 0.7--35~keV \nustar/\swift\ continuum to a model consisting of thermal Comptonization and a soft emission component. For the latter we used \textsc{diskbb}, and for the former we used \textsc{nthcomp} with seed photons provided by a disk black body \citep[][]{zdziarski1996,zycki1999}. The \textsc{nthcomp} model has a power-law component with an index $\Gamma$, a low-energy cutoff set by the seed photon temperature ($kT_{\mathrm{s}}$) and a high energy roll-over determined by the electron temperature ($kT_{\mathrm{e}}$). This two-component model often provides a good description of the broad-band continuum of neutron star LMXBs in their soft spectral states \citep[e.g.,][]{barret2002,piraino2007,disalvo2015,king2016}. It may be physically interpreted as direct emission from the accretion disk and Comptonized emission from either a hot flow or a boundary layer where the disk runs into the neutron star surface, with the seed photons provided by the disk and/or surface of the neutron star \citep[e.g.,][]{barret2000,disalvo2000}.

Our continuum model can describe the 0.7--35~keV \nustar/\swift\ spectral data reasonably well ($\chi^2_{\nu}$$=$1.05 for 1495 dof; Figure~\ref{fig:contspec}). We obtain a hydrogen column density of $N_{\mathrm{H}}$$=$$(3.3\pm 0.2)\times10^{21}~\nh$ and an inner disk temperature of $kT_{\mathrm{in}}$$=$$0.89\pm 0.07$~keV. For the \textsc{nthcomp} component we obtain $\Gamma$$=$$2.6 \pm 0.1$, $kT_{\mathrm{s}}$$=$$1.51\pm 0.13$~keV, and $kT_{\mathrm{e}}$$=$$2.72 \pm 0.07$~keV. The  0.7--35~keV unabsorbed flux inferred from this fit is $F_{\mathrm{0.7-35}}$$\simeq$$4.7\times10^{-9}~\flux$, which translates into a luminosity of $L_{\mathrm{0.7-35}}$$\simeq$$1.9\times10^{37}~\dist~\lum$. The disk contribution to the total 0.7--35 keV flux is $\simeq$35\%. We note that a continuum consisting of a black body, a (broken) power law and a disk black body is also frequently used for soft state spectra of neutron star LMXBs \citep[e.g.,][]{lin2007,cackett2010_iron,miller2013_serx1}. Such a model fits our data less well ($\chi^2_{\nu}$$=$1.06 for 1495 dof). 

Figure~\ref{fig:contspec} shows that our continuum description leaves residuals near 6--7 keV. These are likely due to a broad Fe-K emission line that arises from disk reflection, as was also seen during the hard state early on in the 2015 outburst of \source\ \citep[][]{ludlam2016}.\footnote{An alternative interpretation of these broad features as a mixture of number of narrower emission lines \citep[see e.g.,][for a discussion]{ng2010} can be ruled out for at least some LMXBs \citep[e.g.,][]{chiang2015}.} We therefore continued our spectral analysis by including physical reflection models.

 \begin{figure}
 \begin{center}
	\includegraphics[width=8.5cm]{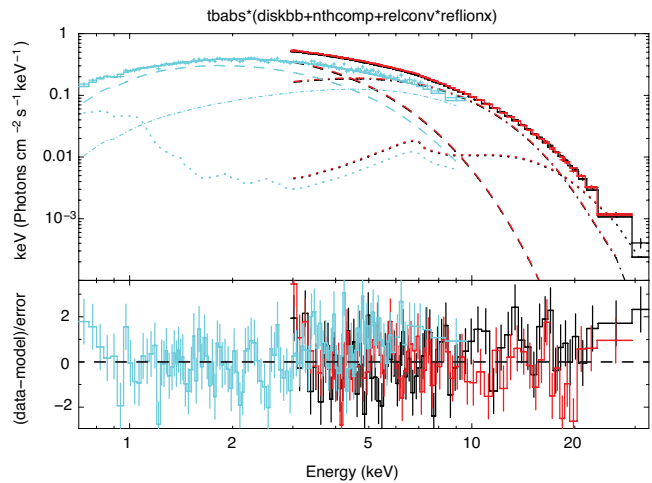}
	    \end{center}
    \caption[]{\nustar\ (FPMA in black, FPMB in red) and \swift/XRT (cyan) unfolded spectra fitted with a  model consisting of a disk black body (dashed curves), Comptonized emission (dashed-dotted curves), and relativistically blurred reflection (dotted curves). The bottom panel shows the residuals in sigmas. The spectral data were rebinned for visual clarity.
        }
 \label{fig:reflspec}
\end{figure}

\begin{table}
\caption{Results from fitting \nustar\ and \swift\ spectral data to a model \textsc{constant*tbabs*(diskbb+nthcomp+relconv*reflionx)}.
}
\begin{threeparttable}
\begin{tabular*}{0.48\textwidth}{@{\extracolsep{\fill}} l c}
\hline 
Parameter (unit) & value \\
\hline
% \multicolumn{2}{c}{  \textsc{constant}  }  \\
$C$  (FMPB) \dotfill & $1.034 \pm 0.002$  \\
$C$  (XRT) \dotfill & $0.682 \pm 0.005$   \\
 &     \\
%  \multicolumn{2}{c}{  \textsc{tbabs}  }  \\
$N_H$  ($\times10^{21}~\nh$) \dotfill & $4.8 \pm 0.2$  \\
 &     \\
% \multicolumn{2}{c}{  \textsc{diskbb}  }  \\
$kT_{\mathrm{in}}$  (keV) \dotfill & $1.15 \pm 0.02 $  \\
$N_{\mathrm{diskbb}}$  $(\mathrm{km}/10~\mathrm{kpc})^2 \cos i$) \dotfill & $91.9^{+5.8}_{-9.1}$  \\
  &     \\
% \multicolumn{2}{c}{  \textsc{nthcomp}  }  \\
$kT_{\mathrm{s}}$  (keV)  \dotfill&  $1.27\pm0.45 $  \\
$kT_{\mathrm{e}}$  (keV) \dotfill&   $1.87\pm 0.04$ \\
$\Gamma$  \dotfill  &   $1.3 \pm 0.2$ \\
$N_{\mathrm{nthcomp}}$  ($\times10^{-2}$) \dotfill&  $9.1\pm0.1$ \\
  &      \\
% \multicolumn{2}{c}{  \textsc{relconv*relfionx}  }  \\
$i$  ($^{\circ}$) \dotfill & $32.2 \pm 3.0$  \\
$R_{\mathrm{in}}$  ($\times \risco$) \dotfill & $1.0^{+0.5}$ \\
 $\xi$  ($\mathrm{erg~cm~s}^{-1}$) \dotfill & $119.4\pm6.8$  \\
 $A_{\mathrm{Fe}}$  ($\times$Solar) \dotfill & $1.2\pm0.4$ \\
 $kT_{\mathrm{refl}}$  (keV) \dotfill &  $2.96 \pm 0.07$ \\
$N_{\mathrm{refl}}$ \dotfill &  $1.8 \pm 0.5$ \\
 &      \\
%  \multicolumn{2}{c}{  \textsc{cflux}  }  \\
$F_{\mathrm{0.7-35}}$ ($\times10^{-9}~\flux$) \dotfill &  $4.80\pm 0.01$  \\
$L_{\mathrm{0.7-35}}$  ($\times 10^{37}~\dist~\lum$) \dotfill &  $1.93\pm 0.01$  \\
 &      \\
$\chi^2_{\nu}$ (dof) \dotfill  &  0.98 (1489)\\
\hline
\end{tabular*}
\label{tab:reflspec}
\begin{tablenotes}
\item[]
Note. -- We fixed $a$$=$0.3, $q$$=$3 and $R_{\mathrm{out}}$$=$$500~\risco$ for \textsc{relconv} and assumed disk black body seed photons for \textsc{nthcomp}. The constant multiplication factor was fixed to $C$$=$1 for the \nustar\ FMPA data, and left free for the other instruments. $F_{\mathrm{0.7-35}}$ gives the total unabsorbed model flux in the 0.7--35 keV band, and $L_{\mathrm{0.7-35}}$ is the corresponding luminosity for $D$$=$5.8~kpc. Quoted errors reflect 1$\sigma$ confidence levels.
\end{tablenotes}
\end{threeparttable}
\end{table}

\subsection{Reflection spectrum}\label{subsec:refl}
For neutron star LMXBs in their soft states, the X-ray spectrum above $\simeq$5~keV is typically modeled as either a hot ($\simeq$2--3~keV) black body or thermal Comptonization. This emission component provides most of the hard X-ray flux that illuminates the accretion disk and produces the reflection spectrum \citep[e.g.,][]{cackett2010_iron}. We therefore used a modified version of \textsc{reflionx} \citep[][]{ross2005} that assumes a black body input spectrum.\footnote{Available at http://www-xray.ast.cam.ac.uk/$\sim$mlparker/reflionx$\_$models/ reflionx$\_$alking.mod.} Although our broad-band fits prefer a Comptonized model over a black body to describe the spectrum above $\simeq$5~keV (see Section~\ref{subsec:broad}), we found that this reflection model works very well for our data. This is presumably because the Comptonized emission can be approximated by a hot black body and because the Fe-K line in our data is relatively weak, as is typically the case in soft spectral states. The fit parameters of the \textsc{reflionx} model are the ionization parameter $\xi$, the iron abundance $A_{\mathrm{Fe}}$ (with respect to Solar), the temperature of the ionizing black body flux ($kT_{\mathrm{refl}}$), and a normalization ($N_{\mathrm{refl}}$). 

To account for broadening due to relativistic effects in the vicinity of the neutron star, we use the blurring kernel \textsc{relconv} with an unbroken emissivity profile of the form $\epsilon \propto r^{-q}$. The fit parameters of the \textsc{relconv} model are the emissivity index $q$, the inner and outer disk radius $R_{\mathrm{in}}$ and $R_{\mathrm{out}}$ (expressed in $\risco$), the disk inclination ($i$), and the dimensionless spin parameter $a$. We fixed $q$$=$3 \citep[appropriate for neutron star LMXBs, see e.g.,][]{cackett2010_iron}, because this parameter was poorly constrained when left free to vary. We also fixed $R_{\mathrm{out}}$$=$$500~\risco$, since this emissivity profile drops off steeply with increasing radius so that the reflection fits are not sensitive to the outer disc radius. 

The spin period of \source\ is unknown, but neutron stars rotate slowly enough for the spin to have little effect on the metric. For  neutron stars, the dimensionless spin parameter can be approximated as $a$$\simeq$$0.47/P_{\mathrm{ms}}$, where $P_{\mathrm{ms}}$ is the spin period in ms \citep[][]{braje2000}. The fastest known neutron stars spin at $\simeq$1.5~ms \citep[e.g.,][]{galloway06}, i.e., $a$$\simeq$0.3. Using equation 3 from \citet{miller1998}, the innermost stable circular orbit (ISCO) is then located at $R_{\mathrm{isco}}$$=$$5.05~\rg$ ($\simeq$11.2~km), where $\rg$$=$$\gmc$ is the gravitational radius (with $G$ the gravitational constant and $M$ the neutron star mass, which we assume to be $1.5~\Msun$). This is thus only a small shift compared to the Schwarzschild metric ($a$$=$0) for which $R_{\mathrm{isco}}$$=$$6~\rg$ ($\simeq$13.3~km). 

Including relativistic reflection significantly improves our spectral fits ($\chi^2_{\nu}$$=$0.98 for 1489 dof). The best fit parameters for the continuum and reflection spectrum are listed in Table~\ref{tab:reflspec}. Our fits suggest that the inner disk was located close to the neutron star, at $R_{\mathrm{in}}$$=$1--1.5$~\risco$ ($\simeq$5--7.5$~\rg$ or $\simeq$11--17~km), and is viewed at an inclination of $i$$=$$29^{\circ}$--$35^{\circ}$. The obtained Fe abundance is consistent with Solar composition ($A_{\mathrm{Fe}}$$=$1.2$\pm0.4$) and the disk appears to be mildly ionized ($\log \xi$$\simeq$2.1). We note that fits performed with $a$=$0$ yielded similar results as for $a$=$0.3$, as expected. Figure~\ref{fig:reflspec} shows our fit results with relativistically blurred reflection included.

The normalization of the \textsc{diskbb} spectral component can formally also be used to probe the inner radial extent of the accretion disk, as it is defined as $N_{\mathrm{diskbb}}$$=$$(R_{\mathrm{in,diskbb}}/D_{10})^2 \cos i$ (with $R_{\mathrm{in,diskbb}}$ in km and $D_{10}$ the distance in units of 10 kpc). However, several (not well-determined) correction factors need to be taken into account to infer inner-disk radii from \textsc{diskbb} fits, e.g., for inner boundary assumptions \citep[][]{gierlinski1999} and spectral hardening \citep[][]{merloni2000}, which may add up to a factor of $\simeq$1.7 \citep[e.g.,][]{kubota2001,reynolds2013}. Furthermore, uncertainties in the absolute flux calibration of the instrument or the spectral model (e.g., the column density or inclination) directly affect the inner disk radius inferred from the \textsc{diskbb} model. 

Bearing these caveats in mind, we note that the measured disk normalization of our best fit (Table~\ref{tab:reflspec}) translates into an inner disk radius of $R_{\mathrm{in,diskbb}}$$=$$6.0\pm 0.4$~km for an inclination of $i$$=$$29^{\circ}$--$35^{\circ}$ (and $D_{10}$$=$0.58), without applying any corrections. Taking into account that the true inner disk radius may be a factor $\simeq$1.7 higher yields $R_{\mathrm{in,diskbb}}$$\simeq$9.5--11~km. This is consistent with the location of the inner disk inferred from our reflection fits ($R_{\mathrm{in}}$$\simeq$11--17~km).

\subsection{High-energy absorption features}\label{subsec:wind}

\subsubsection{\swift/XRT soft-state spectra}\label{subsec:xrtabs}
\begin{figure}
 \begin{center}
	\includegraphics[width=8.3cm]{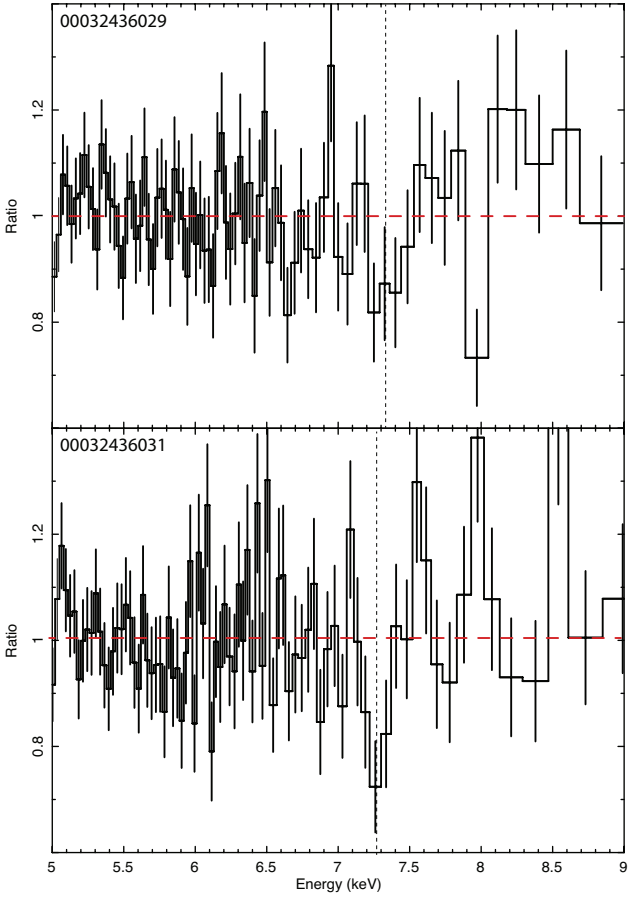}
    \end{center}
    \caption[]{\swift/XRT data to model ratio in the 5--9 keV range for a simple 2-component continuum consisting of a black body and a disk black body. Negative residuals are present near $\simeq$7.2~keV (dotted vertical lines), which were not seen during the other three soft-state observations.
        }
 \label{fig:swiftspec}
\end{figure}

\citet{degenaar2015_1804} reported hints of narrow absorption features near 7--8~keV in \swift/XRT spectra obtained after the transition to the soft state. This is illustrated by the top panel in Figure~\ref{fig:swiftspec}, which shows the high-energy residuals after fitting the first soft-state observation (ID 32436029) to a simple two-component continuum consisting of a black body and a disk black body.\footnote{When using \textsc{diskbb+nthcomp}, several of the parameters were not well-constrained, presumably due to the narrow energy range and limited sensitivity of the XRT data. We therefore used this simpler continuum model for the \swift\ data, which provided a good description.} 

Adding a Gaussian (\textsc{gauss}) with a central energy constrained to 7--8.5~keV suggests that an absorption feature may be present at $E_{l}$$=$$(7.26 \pm 0.06)$~keV. The width of the line appears unresolved and was therefore fixed to $\sigma_{l}=10^{-4}$~keV. Dividing the normalization of the line by its 1$\sigma$ error suggests that the feature is $\simeq$3.2$\sigma$ significant. We measure an equivalent width of $EW$$\simeq$82~eV.
As can be seen in the bottom panel of Figure~\ref{fig:swiftspec}, a similar feature is seen at roughly the same energy in an observation performed 5 days later (ID 32436031; see Table~\ref{tab:obs}). Adding a Gaussian to the continuum fit for this observation returns a central line energy of $E_{l}$$=$$(7.18 \pm 0.05)$~keV, a significance of $\simeq$3.6$\sigma$, and $EW$$\simeq$89~eV. We note that considering the number of trials by counting the observations searched (5) and the number of resolution elements in the search area (7), the significance of the features reduces to $\simeq$2$\sigma$ and $\simeq$3$\sigma$ for observation 32436029 and 32436031, respectively. The line parameters and the continuum fluxes are listed in Table~\ref{tab:speclines}.

Since the line energies obtained from the two spectra have overlapping 1$\sigma$ errors and the measured continuum fluxes were similar (see Table~\ref{tab:speclines}), we also fitted them together with the line parameters tied. This yielded $E_{l}= (7.24 \pm 0.02)$~keV and a significance of $\simeq$4.8$\sigma$ ($\simeq$4$\sigma$ after accounting for trials). The 0.5--10 keV luminosity during these two observations was $L_{0.5-10}$$\simeq$$(2.9-3.2) \times 10^{37}~\dist~\lum$. There are no features of similar significance detected in the other three soft-state XRT observations.

\subsubsection{Narrow absorption lines in the \chan/HEG data}\label{subsec:chanabs}
The \chan/HEG data provide superior spectral resolution and are therefore more suitable to detect and characterize narrow absorption features. Our observation was performed $\simeq$9 days after the transition to the soft state, and $\simeq$5--9 days after the possible absorption line detections in the \swift/XRT data (see Table~\ref{tab:obs}).

\begin{figure}
 \begin{center}
 	\includegraphics[width=8.3cm]{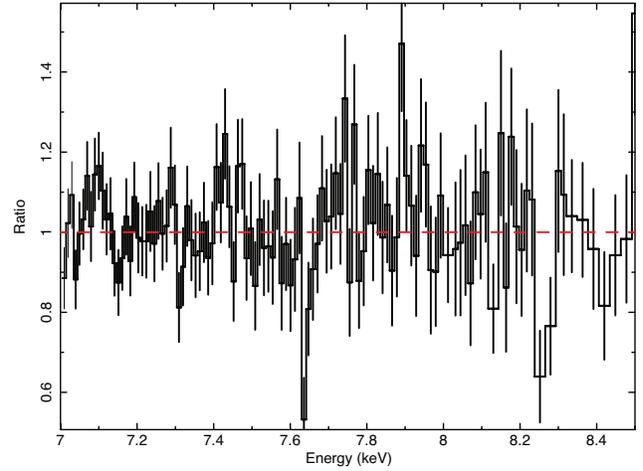}\vspace{+0.2cm}
   \end{center}
    \caption[]{
    Data to model ratio in the \chan/HEG data zoomed at the 7--8.5 keV range. Narrow absorption features are seen near $\simeq$7.6 and 8.2~keV. 
        }
 \label{fig:chanspec}
\end{figure}

To have the best constraints on the continuum, we fitted the \chan\ data together with our simultaneous \nustar\ and \swift\ observations to our model described in Section~\ref{subsec:refl} (Table~\ref{tab:reflspec}). This fit leaves negative residuals in the HEG data near $\simeq$7 and $\simeq$8~keV, as shown in Figure~\ref{fig:chanspec}. We characterized these two absorption features by adding simple Gaussian lines to our spectral fits and estimated their significance by dividing the normalization of the lines by the 1$\sigma$ errors. The line normalization was set to zero for the \nustar\ and \swift\ data because the energy resolution of these instruments renders them insensitive to the very narrow features that we find in the \chan\ data (see below). The width of the lines was unconstrained and therefore fixed at $\sigma_l$$=$$10^{-4}$~keV. We also noted an absorption feature in the HEG data near 5.5~keV, but it was not present in the MEG data and hence likely spurious. 

The line properties and continuum flux are listed in Table~\ref{tab:speclines}. For the first line we obtain a central energy of $E_l$$=$$7.64\pm0.01$~keV, a significance of $\simeq$5.8$\sigma$, and an equivalent width of $EW$$\simeq$18~eV. A second absorption feature is found at $E_l$$=$$8.26\pm0.07$~keV, with a significance of $\simeq$3.6$\sigma$ and $EW$$\simeq$23~eV. When accounting for the number of trials, the significance decreases to $\simeq$4.8$\sigma$ and $\simeq$2$\sigma$, respectively. Both features are also seen in the separate plus and minus order spectra, albeit at lower significance. There are no chip gaps near the inferred line energies. However, given the low significance and unclear physical interpretation, we consider it most likely that the 8.26~keV feature is spurious (see Section~\ref{subsubsection:photoion}). Moreover, we point out that there are also several concerns about the physical interpretation of the $\simeq$7.64~keV feature (see Section~\ref{subsec:wind}).

\begin{table}
\caption{Narrow absorption lines in \chan\ and \swift\ spectral data.}
\begin{threeparttable}
\begin{tabular*}{0.48\textwidth}{@{\extracolsep{\fill}} l c c c}
\hline 
Instrument & ObsID & $E_{\mathrm{l}}$ & $F_{0.5-10}$ \\
& & (keV) & ($\flux$) \\
\hline
\swift/XRT & 32436029  & $7.26 \pm 0.06$ & $7.2 \times 10^{-9}$ \\
\swift/XRT & 32436031  & $7.18 \pm 0.05$ &  $7.9 \times 10^{-9}$  \\
\swift/XRT & 32436029+31  & $7.24 \pm 0.04$ &   \\
\chan/HEG & 17649  & $7.64 \pm 0.01$ &  $5.2 \times 10^{-9}$ \\
 &   & $8.26 \pm 0.01$ &    \\
\hline
\end{tabular*}
\label{tab:speclines}
\begin{tablenotes}
\item[]Note. -- $F_{0.5-10}$ gives the unabsorbed 0.5-10 keV continuum flux. The line widths were unconstrained when left to vary and were therefore fixed to $\sigma_l=10^{-4}$. Errors reflect 1$\sigma$ uncertainties.
\end{tablenotes}
\end{threeparttable}
\end{table}

\subsubsection{Photo-ionization modeling of the \chan\ data}\label{subsubsection:photoion}
Assuming that the narrow lines in the \chan/HEG data are real, we briefly explored what physical properties of the absorbing gas would be required to give rise to these features. To this end, we created a grid of self-consistent photoionization models with \textsc{xstar} \citep[][]{kallman2001}. These were read into \textsc{XSpec} as a multiplicative model to fit the HEG spectral data. The \textsc{xstar} modeling can be regarded as a consistency check to determine if it is possible to create the observed absorption lines with reasonable physical parameters, without producing any other obvious features that are not seen.

To create a grid for \source, we extrapolated our continuum fit to an energy range of 0.2--35 keV and supplied this to \textsc{xstar} as the source of ionizing flux with a bolometric luminosity of $L$$=$$5\times10^{37}~\lum$ (inferred from extrapolating our continuum fit to 0.01--100 keV). We used a fiducial covering factor of $\Omega/4\pi$$=$0.5, which is typically a good assumption for disk winds in LMXBs \citep[e.g.,][]{miller2015_winds}. Furthermore, we assumed a number density of $n$$=$$10^{12}~\dens$, a turbulent velocity of $200~\kms$, and Solar abundances as input for the grid calculations. The fit parameters for our \textsc{xstar} model are the matter density $N$, the ionization parameter $\xi$, and the blueshift $v_{out}$. Since these fits are non-unique and are only intended to be exploratory, we do not give errors on the fit parameters. 

The most abundant elements seen in absorption in the X-ray spectra of LMXBs are Fe\,\textsc{xxv} (rest wavelength 6.70~keV) and Fe\,\textsc{xxvi} (rest wavelength 6.97~keV), which can be red/blue shifted in case of an in/outflow \citep[e.g.,][]{ponti2012_winds,diaztrigo2015}. We found that any model fitting the $\simeq$7.64~keV line as blue-shifted Fe\,{\sc xxv} predicts a strong Fe\,{\sc xxvi} line at commensurate blue-shift ($\simeq$7.98~keV), which is not seen. A higher density or lower ionization does not solve this issue, as that results in prominent absorption lines near 2.3 and 3 keV that are not seen either. 

The 7.64 and 8.26~keV features can thus not arise from the same absorbing plasma, as the implied blueshifts would be different. Given this discrepancy and the low significance of the 8.26~keV feature ($\simeq$2$\sigma$ after accounting for the number of trials), we consider it likely that this feature is spurious. If we assume that only the $\simeq$7.64~keV line is real, its most likely interpretation is blue-shifted Fe\,{\sc xxvi} and would thus suggest an outflow. We can then model this feature with $N$$\simeq$$2\times10^{22}~\nh$ and $\log \xi$$\simeq$4.3, which are plausible values for LMXBs. With these gas parameters no other absorption lines should be seen, consistent with our data. The implied outflow velocity is $v_{\mathrm{out}}$$\simeq$$0.086c$$\simeq$$ 25\,800~\kms$. We note that the gas density and ionization parameter obtained for this exploratory fit should not be taken at face value. In particular, the geometry of the putative absorber in \source\ is unclear (see Section~\ref{subsec:wind}) and the covering fraction is degenerate with the column density.

%%%%%%%%%%%%%%%%%
% DISCUSSION
%%%%%%%%%%%%%%%%%

\section{Discussion}\label{sec:discuss}
We report on \nustar, \swift, and \chan\ observations obtained during a $\simeq$11-day window around the peak brightness of the 2015  outburst of the neutron star LMXB \source. The broad-band 0.7--35 keV \nustar/\swift\ spectral data can be described by a continuum model composed of a disk black body and thermal Comptonization. The source was in a soft spectral state during our observations and we measured a 0.5--10 keV luminosity of $L_{0.5-10}$$\simeq$$(2-3) \times 10^{37}~\dist~\lum$. Assuming a bolometric correction factor of 2--3 \citep[e.g.,][]{zand07}, this corresponds to $\simeq$10\%--25\% of the Eddington limit \citep[$L_{\mathrm{Edd}}$$=$$3.8\times10^{38}~\lum$;][]{kuulkers2003}.

Superimposed on the continuum, we detect a broad Fe-K emission line that likely arises from hard X-rays reflecting off the accretion disk. Modeling this feature as relativistically blurred reflection allows us to constrain the accretion geometry in the soft state. We find a disk inclination of $i$$=$29$^{\circ}$--35$^{\circ}$ (1$\sigma$ confidence level), consistent with results obtained from reflection analysis in the hard state of the same \source\ outburst \citep[$i$$=$18$^{\circ}$--29$^{\circ}$ at 90\% confidence;][]{ludlam2016}. 

Our analysis suggests that the inner edge of the accretion disk extended inwards to $R_{\mathrm{in}}$$\simeq$5--7.5$~\rg$ ($\simeq$11--17~km). Bearing in mind the caveats stated in Section~\ref{subsec:refl}, we note that a similar inner disk radius is implied by the normalization of the disk black body component in our spectral fits ($R_{\mathrm{in}}$$\simeq$9.5--11~km). These values are consistent with the expected radius of a neutron star \citep[e.g.,][for a recent overview]{lattimer2014}, which suggests that the accretion disk may have been truncated by the stellar surface (and hence that the ISCO lies within the neutron star). 

The inner disk may have also been truncated by the magnetic field of the neutron star, rather than its surface. We can thus use our measured inner disk radius to estimate an upper limit for the magnetic field strength of the neutron star. To this end we use equation (1) of \citet{cackett2009_iron}, which was adapted from the formulation of \citet{ibragimov2009}. Extrapolating our \nustar/\swift\ spectral fit to 0.01--100 keV suggests a bolometric flux of $F_{\mathrm{bol}}$$\simeq$$1.2\times10^{-8}~\flux$. We assume $D$$=$5.8~kpc, $M$$=$$1.5~\Msun$, $R$$=$10~km and make similar assumptions regarding geometry and the accretion efficiency as in \cite{cackett2009_iron}; an anisotropy correction factor $f_{\mathrm{ang}}$$=$1, a geometry coefficient $k_{\mathrm{A}}$$=$0.5, and an accretion efficiency $\eta$$=$0.1. The constraint that $R_{\mathrm{in}}$$\lesssim$$7.5~Rg$, then yields $B$$\lesssim$$2\times10^{8}$~G for \source.

\subsection{On the possible detection of a disk wind}\label{subsec:wind}
We detect a narrow absorption line at $\simeq$7.64~keV in our \chan/HEG data that is $\simeq$4.8$\sigma$ significant after accounting for the number of trials. Other narrow absorption features were seen in the \chan/HEG data and two of the \swift/XRT soft-state observations at energies of $\simeq$7--8 keV, but at lower significance ($\simeq$2--4$\sigma$ after accounting for trials). Blue-shifted, narrow absorption lines have been seen in several black hole and neutron star LMXBs during their soft states, and are interpreted as outflowing disk winds \citep[e.g.,][]{brandt2000,lee2002,ueda2004,miller2006,miller2008,miller2011,miller2016_wind,diaztrigo2007,neilsen2009,neilsen2011,king2012,king2015,ponti2012_winds,degenaar2014_groj1744,bozzo2016_wind}. 

Since \source\ was in a soft state during the observations analyzed in this work, it is plausible that a disk wind was present. Assuming that the 7.64~keV line in our \chan\ data is both real and due to a disk wind, our photo-ionization modeling suggests that the most likely identification is blue-shifted Fe\,{\sc xxvi}, which would imply an outflow velocity of $\simeq$$0.086c$$\simeq$$ 25\,800~\kms$. However, we note that a solid identification is not trivial with just a single line. Moreover, there are several concerns about interpreting this feature as being due to a fast, outflowing disk wind. 

Firstly, observational evidence suggests that disk winds are concentrated in the accretion disk plane \citep[e.g.,][]{miller2006_winds,miller2006,miller2015_winds,king2012,ponti2012_winds,degenaar2014}. Indeed, simulations of (thermally-driven) disk winds suggest that at near-polar angles, the low density and high ionization of the gas may prevent its detection \citep[e.g.,][]{sim2008,higginbottom2015}. However, our spectral analysis suggests that \source\ is seen at a relatively low inclination of $i$$\simeq$$30^{\circ}$ \citep[see also][]{ludlam2016}. It is therefore not clear whether a disk wind, if present, would be observable. Nevertheless, a signature of a wind was recently reported for the neutron star LMXB GX 340+0, which has a similarly low inclination of $i$$\simeq$$35^{\circ}$ \citep[][]{miller2016_wind}. It was proposed that in this source the wind may be radiatively or magnetically driven \citep[see e.g.,][for a discussion on wind driving mechanisms]{diaztrigo2015}. Indeed, simulations of magnetically-driven winds in LMXBs suggest that these may be observable at lower inclination angles \citep[e.g.,][]{chakravorty2015}.

Secondly, the implied outflow velocity of $\simeq$$25\,800~\kms$ ($\simeq$0.086$c$) would be extreme for an LMXB. Such fast outflows are not uncommon in supermassive black holes \citep[e.g.,][for a recent review]{tombesi2016}, but wind velocities tend to be more modest in LMXBs. The highest velocities claimed for LMXBs so far are $\simeq$$9\,000-14\,000~\kms$ \citep[$\simeq$0.03--0.04$c$;][]{king2012,degenaar2014_groj1744,miller2016_wind}, but more typical values are $\simeq$$400-3\,000~\kms$ \citep[$\simeq$0.001--0.01$c$; e.g.,][for an overview]{diaztrigo2015}. Wind velocities in LMXBs are thus commonly a factor $>$10 lower than implied for \source, and even the most extreme cases so far have a factor $>$3 lower outflow velocities than we find. 

Although there is reason to be skeptical about the detection of the $\simeq$7.64~keV line, we briefly speculate on its possible implication assuming that it is both real and that its interpretation as blue-shifted Fe\,{\sc xxvi} is correct. Observations seem to suggests that winds and jets generally do not co-exists, and leave room for the possibility that winds and jets perhaps evolve into one another \citep[e.g.,][]{miller2006_winds,neilsen2009,king2012,ponti2012_winds,diaztrigo2013_nature}. If so, winds may accelerate and become denser away from the disk plane when moving toward lower luminosity. This could perhaps account for detecting a fast outflow from \source\ in our \chan\ observations, as it was accreting at a relatively low luminosity of $L_{0.5-10}$$\simeq$$2 \times 10^{37}~\dist~\lum$ at the time. Within this context it is interesting to note that \source\ was detected at radio wavelengths on 2015 April 12, just two days before our \chan\ observation, albeit with a much lower flux density than during the preceding hard spectral state (Gusinskaia et al., in preparation). 

It is also possible that the neutron star magnetic field acts as a propeller and ejects material. MHD simulations of \citet{romanova2009} reveal both a jet-like and a wind-like outflow for an active propeller. In these simulations the wind component is a thin conical shell with a half-opening angle of $\simeq$$30^{\circ}$--$40^{\circ}$ and a velocity up to $\simeq$0.1$c$. Such an outflow may be compatible with our X-ray observations of \source\ and the quasi-simultaneous weak radio detection. In particular, the putative X-ray absorption lines are very narrow and it would require a thin ejected shell to be internally consistent with the radial velocity implied by the blue-shift. 

A propeller operates when the magnetospheric radius is larger than the co-rotation radius (i.e., the radius at which the Keplerian frequency of the disk is equal to the spin frequency of the neutron star). If the inner disk in \source\ were to be truncated by the magnetosphere and this radius ($R_{\mathrm{in}}$$\lesssim$17~km) were to be larger than the co-rotation radius, then the neutron star would have to be spinning at $\lesssim$0.4~ms. However, no neutron stars with sub-millisecond spin periods are known to date \citep[e.g.,][]{patruno2012_gravwav}. 

In summary, there is an absorption feature significantly detected at $\simeq$7.64~keV in our \chan/HEG data that can plausibly be identified as blue-shifted Fe\,{\sc xxvi} and arise from a disk wind. However, this would require that winds in neutron star LMXBs are observable at low inclination angles of $i$$\simeq$30$^{\circ}$, and are able to reach an outflow velocity as high as $v_{\mathrm{out}}$$\simeq$$25\,800~\kms$.

\subsection{On the size of the binary orbit}\label{subsec:ucxb}
\citet{baglio2016} reported that an optical spectrum obtained  in late April, after \source\ had transitioned to the soft X-ray spectral state, was featureless apart from a possible He{\sc ii} emission line at 4686~\AA. Based on the lack of H-emission lines, which are usually seen in the optical spectra of LMXBs, the authors proposed that the donor star must be H poor and thus that \source\ is a candidate ultra-compact X-ray binary (UCXB). UCXBs are X-ray binaries in which the donor is an evolved star in a tight $P_{\mathrm{orb}}$$\lesssim$90~min orbit. \citet{baglio2016} estimated the time-averaged mass-accretion rate of \source\ based on \rxte/ASM and \maxi\ monitoring, and concluded that this is broadly consistent with the source having a He-rich donor in a $\simeq$40~min orbit, based on a comparison with the disk instability model (\citet{lasota2008}; see also the evolutionary tracks of \citet{vanhaaften2012}).

We note that the lack of H-emission lines during single-epoch spectroscopy does not necessarily has to imply an UCXB nature. There are a few (black hole) LMXBs that did not show H-emission lines in their optical outburst spectra at some epochs, but have measured orbital periods of $P_{\mathrm{orb}}$$\gtrsim$90~min: e.g., Swift J1357.2--0933 \citep[][]{torres2011}, and Swift J1753.5--0127, \citep[][]{jonker2008_faint}. Nevertheless, their orbits are relatively small: $P_{\mathrm{orb}}$$=$2.8 and 2.85 or 3.2~hr, respectively \citep[][]{zurita2008,corralsantana2013,neustroev2014}. This may suggests a possible link between a short orbital period and the absence of H-emission lines. Indeed, there are several additional arguments that would support a relatively short orbital period for \source.

Firstly, the full-width half maximum (FWHM) of the tentative He{\sc ii} 4686-\AA\ line identified by \citet{baglio2016} is quite low. From their figure~1, we estimate a FWHM of $\simeq$15~\AA\ and hence an equivalent resolved velocity of $\simeq$960$~\kms$. If a similar FWHM-$K_2$ correlation as recently found by \citet{casares2015} would apply for He and during outburst, this would suggest a low velocity amplitude of the companion star of $K_2$$=$$0.233\times$FWHM$\simeq$224$~\kms$. For a short orbital period of 1--3 hr and a very small mass ratio (applicable for a short period system), the mass function then is $f$$\simeq$0.05--0.14~$\Msun$. For a 1.5~$\Msun$ neutron star, this implies a low inclination of $i$$\simeq$19$^{\circ}$--27$^{\circ}$, which is remarkably similar to that obtained from modeling the Fe-K line in the X-ray spectra \citep[$i$$\simeq$18$^{\circ}$--35$^{\circ}$; this work and][]{ludlam2016}. Such a low inclination could account for the lack of orbital signatures in the optical data \citep[][]{baglio2016}, as ellipsoidal modulations would be small. 

Secondly, the orbital period of LMXBs correlates with their absolute visual magnitude and X-ray luminosity (\cite{vanparadijs94}, see also \cite{russell2006}). We measured an X-ray luminosity of $L_{0.7-35}$$\simeq$$2\times10^{37}~\lum$ on 2015 April 14, whereas \citet{baglio2016} reported an apparent visual magnitude of $\simeq$17~mag on 2015 April 24 (see their table 2). For $D$$=$5.8~kpc, we can then roughly estimate an orbital period of $P_{\mathrm{orb}}$$=$2.7~hr using the empirical relation of \citet{vanparadijs94}.

The X-ray spectral shape may also provide clues about the size of the binary. \citet{sidoli2001} analyzed the broad-band (0.1--100 keV) spectra of a sample of Galactic globular cluster LMXBs, fitting these with a disk black body and a Comptonizing component. These authors found that only in the (candidate) UCXBs, the obtained inner disk temperatures were similar to that of the seed photons of the Comptonized emission, and the obtained radii (several km) broadly consistent with that expected for the location of the inner disk. For the non-UCXBs, the inner disk temperatures were much higher ($\simeq$2--3~keV), and their radii un-physically small ($\lesssim$1~km). This seems to suggest that only in UCXBs, the region providing the seed photons of the Comptonized emission is small and consistent with the inner accretion disk \citep[][]{sidoli2001}. 

The tentative classification scheme of \citet{sidoli2001} gained credence through testing for an UCXB nature by alternative methods such as the relation of \citet{vanparadijs94} discussed above, and the composition of the accreted matter inferred from thermonuclear X-ray bursts \citep[see][]{verbunt2006}. Its validity has also been strengthened by a number of additional sources \citep[e.g.,][]{gierlinski2005,falanga2005,fiocchi2008}. However, \citet{engel2012} identified a 2.15-hr orbital period for XB~1832--330 in the Galactic globular cluster NGC 6652, which was originally put forward as an UCXB based on the spectral-shape classification \citep[][]{parmar2001}. It was therefore proposed by \citet{stacey2012}, that the \citet{sidoli2001} scheme may not strictly apply to UCXBs (i.e., with $P_{\mathrm{orb}}$$\lesssim$90~min), but can still distinguish between short and long orbital period systems.

Against this background, we note that our fits of the 0.7--35 keV \swift/\nustar\ spectrum of \source\ yield an inner disk temperature and seed photon temperature of the Comptonized component that are consistent within their 1$\sigma$ errors; $kT_{\mathrm{in}}$$=$$1.15\pm0.02$~keV and $kT_{\mathrm{s}}$$=$$1.27\pm0.45$~keV, respectively. Furthermore, from the continuum spectral shape we obtain an inner disk radius of $\simeq$9.5--11~km if we boldly apply a color-correction factor of 1.7 (see Section~\ref{subsec:refl} for caveats), i.e., consistent with inner disk radius measured from the reflection component. The \citet{sidoli2001} classification scheme would thus favor a small disk size, hence small orbital period, for \source.

Finally, the maximum luminosity that can be reached during outburst should scale with the orbital period of the binary \citep[e.g.,][]{lasota01}, which is borne out by observations \citep[e.g.,][]{wu2010}.
For the estimated bolometric peak X-ray luminosity of \source\ ($L_{\mathrm{bol}}$$\simeq$$5\times10^{37}~\lum$), the empirical relation of \citet{wu2010} yields an orbital period of $\simeq$1.4~hr. Disk instability theory predicts an orbital period of $\simeq$2.4~hr for an irradiated He disk \citep[][]{lasota2008_aa}. A He-dominated disk would be consistent the lack of H-features and presence of He{\sc ii} emission in the optical spectra \citep[][]{baglio2016}. We note that if \source\ harbors an evolved companion, the detection of Fe-K reflection in both the soft and the hard X-ray spectral state might argue in favor of a He donor rather than a C/O or O/Ne/Mg white dwarf donor, since a high C/O abundance could screen Fe and hence suppress the Fe-K line in C/O rich systems (\cite{koliopanos2013}, \cite{koliopanos2014}, but see \cite{madej2014}). 

In conclusion, the lack of H and the narrowness of the tentative He line in the optical spectrum, the X-ray and optical luminosity, the broad-band X-ray spectral shape, and the peak X-ray luminosity are all consistent with \source\ being a neutron star LMXB with a relatively small orbit of $P_{\mathrm{orb}}$$\lesssim$3~hr.

\section*{Acknowledgements}
We are grateful to Belinda Wilkes, Fiona Harrison, Neil Gehrels and the \chan, \nustar\ and \swift\ teams for carrying out these DDT observations. The authors thank the anonymous referee for constructive comments. N.D. thanks Ciro Pinto, Matt Middleton and Anne Lohfink for useful discussions. N.D. is supported by an NWO/Vidi grant and an EU Marie Curie Intra-European fellowship under contract no. FP-PEOPLE-2013-IEF-627148. D.A. acknowledges support from the Royal Society. J.C.A.M.J. is supported by an Australian Research Council (ARC) Future Fellowship (FT140101082) and an ARC Discovery Grant (DP120102393). R.W. and A.P. are supported by an NWO/TOP grant, module 1, awarded to R.W. C.O.H. is supported by an NSERC Discovery Grant. A.T.D. is supported by an NWO/Veni grant. J.W.T.H. is supported by NWO/Vidi and ERC/starting (337062) grants.

\end{document}